\documentclass[twocolumn]{aa}
\usepackage{epsfig}
\def\simlt{\ \raise -2.truept\hbox{\rlap{\hbox{$\sim$}}\raise5.truept   %
\hbox{$<$}\ }}
\def\simgt{\ \raise -2.truept\hbox{\rlap{\hbox{$\sim$}}\raise5.truept   %
\hbox{$>$}\ }}                                                          %

\def\be{\begin{equation}}
\def\ee{\end{equation}}
\def\newline{\hfil\break}

\def\la{\mathrel{\hbox{\rlap{\hbox{\lower4pt\hbox{$\sim$}}}\hbox{$<$}}}}
\def\ga{\mathrel{\hbox{\rlap{\hbox{\lower4pt\hbox{$\sim$}}}\hbox{$>$}}}}

\begin{document}

\title{The SZ effect from cluster cavities}

   \author{S. Colafrancesco }

   \offprints{S. Colafrancesco}

\institute{   INAF - Osservatorio Astronomico di Roma,
              via Frascati 33, I-00040 Monteporzio, Italy.\\
              Email: cola@mporzio.astro.it
             }

\date{Received: January 20, 2005 / Accepted: March 19, 2005}

\authorrunning {S. Colafrancesco}

\titlerunning {The SZ effect from cluster cavities}

\abstract{In this Letter we derive the SZ effect (SZE) observable in the direction of the giant X-ray
cavities found in the atmospheres of galaxy clusters. We explore specifically the case of the cluster
MS0735.6+7421. The SZE produced in the cavities embedded in the cluster thermal atmosphere is of
non-thermal nature: it dominates the overall cluster SZE at the frequency $x_{\rm 0,th}$ -- at which
the thermal SZE from the cluster vanishes -- and appears, at sub-mm frequencies, as a negative SZE
located only in the regions of the intracluster space occupied by the cavities. We show that SZ
observations, combined with X-ray and radio data, are able to shed light on the morphology, on the
physical structure, on the dynamics and the origin of these recently discovered non-thermal features
of galaxy clusters. These studies will be also relevant to determine the impact of specific events of
the nature of cavities on the use of SZ and X-ray clusters as probes for cosmology and for the large
scale structure of the universe.

}

 \maketitle


\section{Introduction}

Cavities with diameters ranging from a few to a few hundreds of kpc have been observed by Chandra in
the X-ray emission maps of several galaxy clusters and groups (see, e.g., Birzan et al. 2004, McNamara
et al. 2005). These cavities are supposed to contain high-energy plasma with a non-thermal spectrum
and are currently studied by combining X-ray and radio observations of galaxy clusters.
The enthalpy (free energy) of such cavities should be of the order of $PV \sim 10^{55} - 10^{60}$ erg
and seems to scale with the cooling X-ray luminosity and with the radio power of the host cluster
(Birzan et al. 2004).
Some specific systems show even the existence of giant cavities, as it is the case of the cluster
MS0735.6+7421 (McNamara et al. 2005), with radii of size $r_{\rm cavity} \sim 100$ kpc which produce a
work $PV \sim 10^{61}$ erg to balance each cavity against the average surrounding pressure $P \approx
6 \times 10^{-11}$ erg m$^{-3}$ of the thermal intracluster (IC) gas. The hot regions found around
these cavities indicate that the IC gas near the cavities is being heated by the shocks associated to
the edge of the cavities. The cavities seem to be filled by the radio lobes of the central radio
source suggesting that the IC gas is displaced and compressed by the advancing radio-emitting plasma.
The plasma contained in the advancing radio lobes is likely to be relativistic with a power-law
electron spectrum $n_{\rm e,rel} \propto E_e^{-\alpha}$ with a typical index $\alpha \sim 2.5$ (see,
e.g., Longair 1993).

While the properties of these cavities and of the relativistic plasma they contain is usually studied
by combining high-resolution X-ray and radio maps, we explore here, as an alternative strategy, the
consequences of the Compton scattering between the high-energy electrons filling the cavities and the
CMB photon field. Such a scattering produces a relativistic non-thermal Sunyaev-Zel'dovich effect
(SZE) (see Sunyaev \& Zel'dovich 1972, 1980; Colafrancesco et al. 2003) whose amplitude, spectral and
spatial features depend on the overall pressure and energetics of the relativistic plasma in the
cavities.
A general approach to the SZE produced by plasma bubbles observed in several clusters has been
recently presented by Pfrommer et al. (2005) who investigated the role of physically different
scenarios for the composition of the plasma bubbles. They suggested - in this context - that there is
a realistic chance to detect an SZ flux decrement from bubbles dominated by relativistic particles.\\
We will show in the following that detailed SZ observations of giant cavities can provide important,
unique (and complementary to X-rays) constraints to the physics and to the evolution of the cavities
in the atmospheres of groups and clusters of galaxies.
We focus here on the specific case of the giant cavities found in the cluster MS0735.6+7421, for which
the available information on the cavity's dimension and internal pressure provide realistic estimates
of the associated SZE spectral and spatial signals.
The relevant physical quantities are calculated using $H_0 = 70$ km s$^{-1}$ Mpc$^{-1}$ and a flat,
vacuum-dominated CDM  ($\Omega_{\rm m} = 0.3, \Omega_{\Lambda}=0.7$) cosmological model.

\section{The SZ effect produced from cluster cavities}

We compute here the spectral and spatial features of the overall SZE seen along the line of sight
(hereafter los) to a cluster cavity. This is given by the combination of the thermal SZE from the IC
ambient medium and the SZE from the cavity plasma. A possible kinematic SZE acts here as a source of
systematic bias (see, e.g., Colafrancesco et al. 2003, Pfrommer et al. 2005).

\subsection{The SZ effect from a cluster cavity}

The generalized expression for the SZ  effect  which is valid in the Thomson limit
for a generic electron population in the relativistic limit and includes also the effects of multiple
scatterings and the combination with other electron population in the cluster atmospheres has been
derived by Colafrancesco et al. (2003) and we will refer to this paper for technical details.
According to these results, the induced spectral distortion observable in the direction of the cluster
cavity can be written as
 \begin{equation}
\Delta I_{\rm cavity}(x)=2\frac{(k_{\rm B} T_0)^3}{(hc)^2}y_{\rm cavity} ~\tilde{g}(x) ~,
\end{equation}
where $T_0$ is the CMB temperature and the Comptonization parameter $y_{\rm cavity}$ is given by
\begin{equation}
y_{\rm cavity}=\frac{\sigma_T}{m_{\rm e} c^2}\int P_{\rm cavity} d\ell ~,
\end{equation}
in terms of the pressure $P_{\rm cavity}$ contributed by the electrons present in the cluster
cavities.
The function $\tilde{g}(x)$, with $x \equiv h \nu / k_{\rm B} T_0$, for the electron population in the
cavity can be written as
\begin{equation}
\label{gnontermesatta} \tilde{g}(x)=\frac{m_{\rm e} c^2}{\langle k_{\rm B} T_{\rm e} \rangle} \left\{
\frac{1}{\tau} \left[\int_{-\infty}^{+\infty} i_0(xe^{-s}) P(s) ds- i_0(x)\right] \right\}
\end{equation}
in terms of the photon redistribution function $P(s)$ and of  $i_0(x) = 2 (k_{\rm B} T_0)^3 / (h c)^2
\times x^3/(e^x -1)$, where we define the quantity
\begin{equation}
 \langle k_{\rm B} T_{\rm e} \rangle  \equiv  \frac{\sigma_{\rm T}}{\tau}\int P d\ell
= \int_0^\infty dp f_{\rm e}(p) \frac{1}{3} p v(p) m_{\rm e} c
 \label{temp.media}
\end{equation}
(see Colafrancesco et al. 2003, Pfrommer et al. 2005) which is the analogous of the average
temperature for a thermal population (for a thermal electron distribution $\langle k_{\rm B} T_{\rm e}
\rangle = k_{\rm B} T_{\rm e}$ obtains, in fact). The photon redistribution function $P(s)= \int dp
f_{\rm e}(p) P_{\rm s}(s;p)$ with $s = \ln(\nu'/\nu)$, in terms of the CMB photon frequency increase
factor $\nu' / \nu = {4 \over 3} \gamma^2 - {1 \over 3}$, depends on the momentum ($p$ normalized to
$m_ec$) distribution, $f_{\rm e}(p)$ of the electrons which are filling the cavity.

We describe here the relativistic electronic plasma within the cavity by a single power-law electron
population with the momentum spectrum
\begin{equation}
 \label{leggep1}
f_{\rm e,rel}(p;p_1,p_2,\alpha)=A(p_1,p_2,\alpha) p^{-\alpha} ~; \qquad p_1 \leq p \leq p_2
\end{equation}
where the normalization term $A(p_1,p_2,\alpha)$ is given by
\begin{equation}
 \label{normal1}
A(p_1,p_2,\alpha) = \frac{(\alpha-1)} {p_1^{1-\alpha}-p_2^{1-\alpha}} ~,
\end{equation}
with $\alpha \approx 2.5$ (analogous results hold for a generic value of $\alpha$). Such an
approximation seems reasonable since the cavities are likely produced by the expanding radio jets of
the central radio source.
In the calculation of the non-thermal SZE, the relevant momentum is  the minimum momentum, $p_1$ of
the electron distribution (which is not constrained by the available observations) while the specific
value of $p_2 \gg 1$ is irrelevant for power-law indices $\alpha > 2$ which are indicated by the
electron spectra observed in radio galaxies.
The value of $p_1$ sets the value of the electron density $n_{\rm e,rel}$ as well as the value of the
other relevant quantities which depend on it, namely the optical depth $\tau_{\rm cavity}$ and the
pressure $P_{\rm cavity}$ of the non-thermal population.
In particular, the pressure $P_{\rm cavity}$, for the case of an electron distribution as in
eq.(\ref{leggep1}), is given by
\begin{eqnarray}\label{press_rel}
 P_{\rm cavity}&=&n_{\rm e,rel} \int_0^\infty dp f_e(p) \frac{1}{3} p v(p) m_e c \\
  & =& \frac{n_{\rm e,rel} m_e c^2 (\alpha
  -1)}{6[p^{1-\alpha}]_{\rm p_2}^{p_1}}
  \left[B_{\frac{1}{1+p^2}}\left(\frac{\alpha-2}{2},
   \frac{3-\alpha}{2}\right)\right]_{\rm p_2}^{p_1} \nonumber
\end{eqnarray}
(see, e.g., Colafrancesco et al. 2003, Pfrommer et al. 2005), where $B_x(a,b)=\int_0^x t^{a-1}
(1-t)^{b-1} dt$ is the incomplete Beta function (see, e.g., Abramowitz \& Stegun 1965). For an
electron population with a double power-law (or more complex) spectrum, analogous results can be
obtained (see Colafrancesco et al. 2003 for details).
Notice that an estimate of the pressure of the non-thermal electrons within the cavity -- as obtained
by X-ray observations (e.g., McNamara et al. 2005) --  yields directly an estimate of the non-thermal
electron density $n_{\rm e,rel}(p_1)$ from eq.(\ref{press_rel}), given the value of $\alpha$ and of
the lower momentum $p_1$ of the electron spectrum.
The optical depth of the non-thermal electron population within the cavity is then given by
\begin{equation}
 \label{tau_p1}
 \tau_{\rm cavity}(p_1) = \sigma_T \int d \ell n_{\rm e,rel}(p_1)
\end{equation}
which takes the value $\tau_{\rm cavity}(p_1) \approx 4.1 \times 10^{-7} [n_{\rm e,rel}(p_1)/10^{-6}
{\rm cm}^{-3}] (r_{\rm cavity}/100 {\rm kpc})$ along the los to the cavity center.

\subsection{The SZE from a cavity embedded in the IC gas}

\begin{figure}[tbp]
\begin{center}
 \epsfig{file=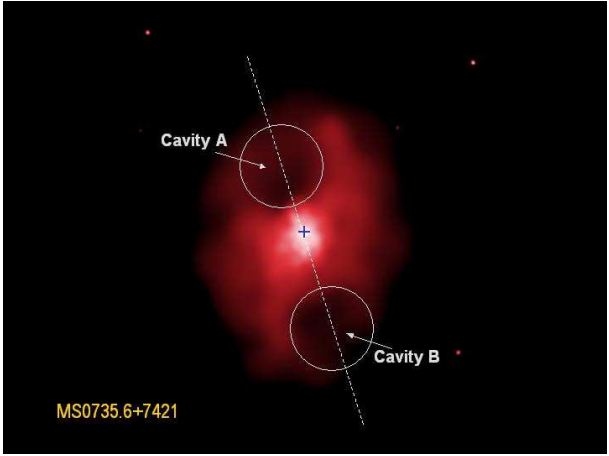,height=6.cm,width=8.cm,angle=0.0}
  \caption{\footnotesize{We show the geometry of the cavities in the cluster MS0735.6+7421
  which has been used to evaluate the SZE. The X-ray image of the cluster MS0735.6+7421 has
  been taken from http://chandra.harvard.edu/photo/2005/ms0735 (credit NASA/CXC/Ohio U./B.McNamara).
  The two cavities have a radius $\approx 100$ kpc and are located at a distance of $\approx 125$ kpc
  and $\approx 170$ kpc from the central radio galaxy (whose position is indicated by a cross)
  along the axis represented in the picture.
  }}\label{fig.sz_cluster}
\end{center}
\end{figure}

\begin{figure}[tbp]
\begin{center}
 \epsfig{file=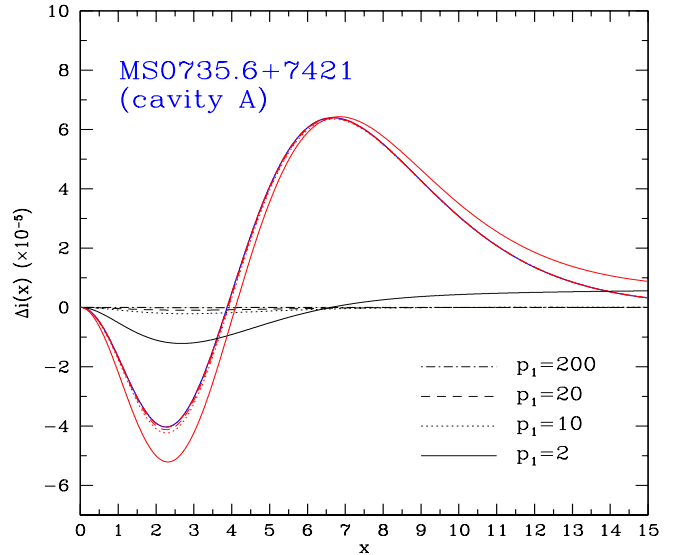,height=8.cm,width=9.cm,angle=0.0}
  \caption{\footnotesize{The spectrum of the overall SZE from the cluster MS0735.6+7421 has been computed
  at a projected radius of $\approx 125$ kpc from the cluster center where the los passes through the
  center of cavity A. We report the thermal SZE (blue), the non-thermal SZE from the
  cavity (black) and the total SZE (red).
  The plotted curves are for different values of the lowest electron momentum: $p_1= 200$
  (dot-dashes), $p_1=20$ (dashes), $p_1=10$ (dots) and $p_1=2$ (solid). The non-thermal SZE is
  normalized to the cavity pressure $P=6 \times 10^{-11}$ erg cm$^{-3}$ and, in this respect
  it must be considered as a lower limit of the true SZE produced in the cavity.
  }}\label{fig.sz_cavity_spec}
\end{center}
\end{figure}

\begin{figure}[tbp]
\begin{center}
 \epsfig{file=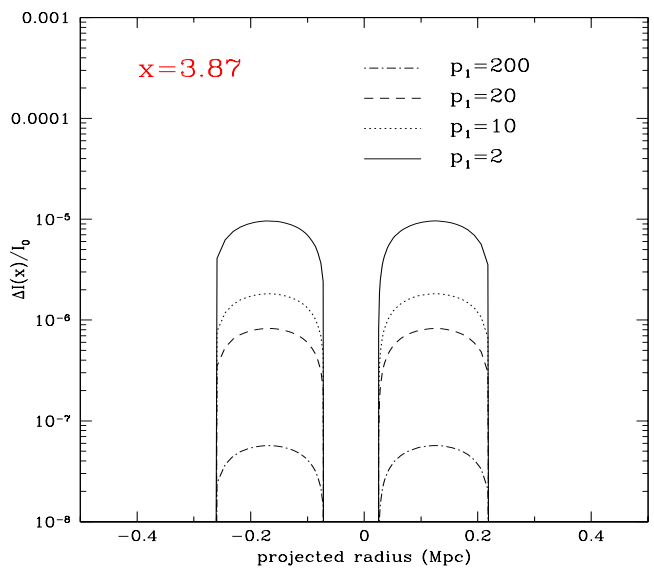,height=8.cm,width=9.cm,angle=0.0}
  \caption{\footnotesize{The spatial distribution of the SZE from the cluster has been computed
  at a frequency $x=3.87$ where the thermal SZE from the cluster MS0735.6+7421 is null.
  At such a frequency, the only visible SZE is that
  emerging from the two cavities A and B of the cluster MS0735.6+7421. A
  possible kinematic SZE could provide a bias of
  $\Delta I / I_0 \approx (1.46 \times 10^{-6}) (\pm v_p/100 {\rm km s}^{-1})$.
  }}\label{fig.sz_cavity_spatial}
\end{center}
\end{figure}

The overall SZE observable along the los through a cluster containing cavities (as shown in
Fig.\ref{fig.sz_cluster} for the representative case of the cluster MS0735.6+7421) is the combination
of the non-thermal SZE produced by the cavity and of the thermal SZE produced by the surrounding IC
gas.
For a proper calculation of the overall SZE, we use the approach described in Colafrancesco et al.
(2003) which we do not repeat here for the sake of brevity. Such a general approach is able to
describe accurately the overall SZE from the combination of the two different electron populations
(non-thermal electrons in the cavity and thermal electrons in the surrounding IC medium) in the full
relativistic limit and with the inclusion of multiple scatterings.
Figs.\ref{fig.sz_cavity_spec} and \ref{fig.sz_cavity_spatial} shows the spectral and spatial features
of the SZE obtained for a case similar to that of the cluster MS0735.6+7421. In our representative and
simplified case, we use a thermal electron distribution with an average temperature $k_{\rm B} T_{\rm
e} = 5$ keV within $70$ arcsec from the cluster center (the region containing the cavities) and an IC
gas density with a radial distribution given by a $\beta$-model with $\beta=1.08$ which reproduces the
thermal pressure profile of the cluster MS0735.6+7421 (see McNamara et al. 2005).
The two cavities are assumed, for simplicity, to be spherical with radius $r_{\rm cavity}=100$ kpc and
with an electron spectrum given by eq.(\ref{leggep1}) with $\alpha  = 2.5$. We also neglect here - for
simplicity - the effect of the shock (local temperature and density boosts) on the SZE, while we will
address the more complex and general case of a cluster with cavities, radial temperature distribution
and finite shock width in a forthcoming paper (Colafrancesco 2005, in preparation).
The non-thermal electron spectrum is normalized to reproduce a constant pressure in the cavity of
$P_{\rm cavity}=6 \times 10^{-11}$ erg cm$^{-3}$, which corresponds to a relativistic electron density
$n_{\rm e,rel} \approx 1.12 \times 10^{-7} {\rm cm}^{-3}$ for a value $p_1 = 2\times 10^2$ and a slope
$\alpha = 2.5$ for the relativistic electron spectrum. Given the value of the pressure $P_{\rm
cavity}$, the relativistic electron density $n_{\rm e,rel}$ decreases for increasing values of $p_1$,
according to eq.(\ref{press_rel}). Because the pressure value $P_{\rm cavity}$ has been obtained from
the properties of the IC gas at the cavity boundaries (see McNamara et al. 2005), it should be likely
considered as a lower limit to the actual pressure within the cavity. As a consequence, the SZE
normalized to the pressure value $P_{\rm cavity}$ should also be considered as a lower limit to the
actual SZE produced by the non-thermal electron distribution within the cavity.
Given the cavity pressure $P_{\rm cavity}$ and the non-thermal electron spectrum slope $\alpha=2.5$,
the relative SZE takes different amplitudes and spectral features for the values $n_{\rm e,rel}$ and
$p_1$ which reproduce the pressure $P_{\rm cavity}$ (see eq.\ref{press_rel} and
Fig.\ref{fig.sz_cavity_spec}).\\
The SZE from the cavities increases at low frequencies $x < 3.87$ [i.e. the frequency $x_{\rm
0,th}(P_{\rm th})$ at which the SZE from the thermal population with pressure $P_{\rm th}$ is zero,
taking into account also the appropriate relativistic corrections -- see Itoh et al. 1998,
Colafrancesco et al. 2003)] for decreasing values of $p_1$ which consistently yield increasing values
of $n_{\rm e,rel}$. This is the frequency range where the non-thermal SZE from the cluster cavities
could be optimally detected (see Fig.\ref{fig.sz_cavity_spec}), since at high frequencies $x> 3.87$,
it is rather low compared to the thermal SZE of the host cluster.\\
The presence of cavities filled with non-thermal electrons also modify the spatial distribution of the
overall SZE from the cluster. In fact, the thermal SZE seen along the los to the cluster (which has a
gas radial profile $\propto [1 + (r/r_c)^2]^{-3 \beta/2 + 1/2}$ as a function of the projected radius
$r$) shows two bumps at the cavity locations for frequencies $x < 3.87$ where the (negative)
non-thermal SZE adds up with the same sign to the (negative) thermal SZE, and shows two depressions
(holes) at the cavity locations for frequencies $x \simgt 3.87$ where the (negative) non-thermal SZE
adds up with the opposite sign to the (positive) thermal SZE. However, at frequencies $x \simgt x_{\rm
0,non-th}$ (which is $\approx 6.5$ for $p_1 \approx 2$) the overall SZE again shows two bumps at the
cavity locations because the (positive) non-thermal SZE adds up with the same sign to the (positive)
thermal SZE which is found at high frequencies.

An interesting feature of the overall SZE is that at the frequency $x_{\rm 0,th}(P_{\rm th})\approx
3.87$ (for a $k_BT = 5$ keV cluster) where the thermal (including relativistic corrections) SZE is
zero, the overall SZE from the cluster is dominated by the non-thermal SZE produced in the cavities:
this is negative in sign, has an amplitude which increases for decreasing values of $p_1$ and is
spatially located only in the cavity regions (see Fig.\ref{fig.sz_cavity_spatial}). Thus, the
observation of the SZE from a cluster containing X-ray cavities at the frequency of the zero of its
thermal SZE provides crucial information directly on the physics of the electron population residing
within the cavities and on their spatial distributions. This fact provides a unique tool to study the
pressure content, the energetics and the spatial distribution of the non-thermal plasma contained in
the cavities without any disturbance by the intervening and surrounding IC medium.

\section{Discussion and conclusions}

We have shown in this Letter that giant X-ray cavities discovered in galaxy clusters can be
potentially and effectively studied by using the SZE produced by the combination of the cluster
thermal atmosphere and of the non-thermal plasma contained in the cluster cavities.
Our results are consistent with the general treatment of the SZE from plasma bubbles described by
Pfrommer et al. (2005). We will emphasize, nonetheless, in the following the main new results of our
work.
We applied our analysis to the specific, reference case of the cluster MS0735.6+7421 which hosts two
giant cavities for which information on the pressure, energetics and enthalpy are available.  These
information allowed us to go beyond the general description level and to make realistic predictions of
the SZE from giant cavities in clusters. We also emphasize that our predictions are likely to be
considered as lower limits to the actual SZE expected from these regions.
At frequencies $x \sim 2.5$ there is the maximum amplitude of the non-thermal SZE from the cavity (for
a given electron spectrum and pressure). This produces a bump in the overall SZE at the cavity
location and the addition of a negative SZE signal to the thermal SZE of the cluster,  a result which
is consistent with the findings of Pfrommer et al. (2005).
At $x \simgt x_{\rm 0,th}(P_{\rm th})$ we have the opposite effect but with smaller amplitudes: a
depression in the SZE at the cavity location and the addition of a negative SZE signal to the positive
thermal SZE.
We emphasize that observations at the frequency $x = x_{\rm 0,th}(P_{\rm th})$ (which is $\approx
3.87$ for a $k_B T=5$ keV cluster) provide a unique tool to probe the overall energetics, the pressure
and the spatial extent of the non-thermal plasma contained in the {giant} cavity, an observation which
is rich in information and complementary to those obtained by X-ray and radio observations of cluster
cavities. At this frequency, in fact, the overall SZE from the cluster reveals only the Compton
scattering of the electrons residing in the cavities without the presence of the intense thermal SZE
observable at lower and higher frequencies. Hence, the SZE from a cluster containing cavities (like
the case of MS0735.6+7421) shows up uncontaminated at frequencies $\sim 220$ GHz: it is less extended
than the overall cluster SZE because it is only emerging from the cavity regions and it is also well
separable because the cavities are well defined in both X-rays and SZ images.
In addition, we also emphasize that the observation of the zero of the non-thermal SZE in the cavities
(which is found at high frequencies, depending on the value of $p_1$ or equivalently on the value
$P_{\rm cavity}$) provides a definite way to determine uniquely the total pressure (see Colafrancesco
et al. 2003) and hence the nature of the electron population within the cavity, an evidence which adds
crucial, complementary information to the X-ray and radio analysis.
A plausible source of bias to these observations could be provided by a possibly relevant kinematic SZ
effect due to the cluster peculiar velocity (see Colafrancesco et al. 2003, Pfrommer et al. 2005 for a
discussion).
However, we have shown that the SZE from the giant cavities in a cluster like MS0735.6+7421 can be
effectively studied at frequency $x = x_{\rm 0,th}(P_{\rm th}) \approx 3.87$ where it is not affected
by the thermal SZE and it is only marginally affected by a possible kinematic SZE even at a level of a
few hundreds km/s (see Fig.\ref{fig.sz_cavity_spatial}).

Our study shows more generally that the combination of high spatial resolution and high sensitivity SZ
observations (even with small field of view) with X-ray and radio data will definitely shed light on
the morphology, on the physical structure, on the dynamics and the origin of these recently discovered
non-thermal features of galaxy clusters. These studies will be also relevant to determine the impact
of specific events of the nature of cavities on the use of SZ and X-ray clusters as probes for
cosmology and for the large scale structure of the universe.

\begin{acknowledgements}
The author thanks the Referee for useful comments. This work is supported by PRIN-MIUR under contract
No.2004027755$\_$003.
\end{acknowledgements}

\end{document}